\documentstyle[11pt]{article}
\let\a=\alpha 
\let\b=\beta 
\let\c=\gamma
\let\C=\Gamma  
\let\d=\delta
\let\e=\epsilon

\let\l=\lambda
\let\L=\Lambda
\let\m=\mu
\let\n=\nu 
\let\r=\rho
\let\s=\sigma

\def\p{\partial}
\def\bd{\begin{document}} 
\def\ed{\end{document}}
\def\be{\begin{equation}}
\def\ee{\end{equation}}
\def\ba{\begin{array}}
\def\ea{\end{array}}
\def\bea{\begin{eqnarray}}
\def\eea{\end{eqnarray}}
\def\nn{\nonumber}
\def\ni{\noindent} 
\let\la=\label 
\let\bl=\bigl 
\let\br=\bigr
\let\Br=\Bigr 
\let\Bl=\Bigl
\let\bm=\bibitem
\def\ft#1#2{{\textstyle{{\scriptstyle #1}
\over {\scriptstyle #2}}}}
\def\fft#1#2{{#1 \over #2}}
\def\sst#1{{\scriptscriptstyle #1}}
\newcommand{\eq}[1]{(\ref{#1})}
\def\eqs#1#2{(\ref{#1}-\ref{#2})}
\def\cites#1#2{\cite{#1}-\cite{#2}}
\def\Hat#1{\widehat{#1}}

\begin{document}

\begin{titlepage}

\hfill{CTP TAMU-16/97}

\hfill{hep-th/9703123}

\hfill{\today}

\vspace{30pt}

\centerline {\Large\bf Super Yang-Mills in $(11,3)$ Dimensions}

\vspace{30pt}

\centerline{{\large Ergin Sezgin}\footnote{Research supported in part by 
NSF Grant  PHY-9411543}}

\vspace{15pt}

\centerline{\it Center for Theoretical Physics, Texas A\&M University,}
\centerline{\it College Station, Texas 77843, U.S.A.}

\vspace{50pt}

\centerline{ABSTRACT}

\vspace{15pt}

A supersymmetric Yang-Mills system in $(11,3)$ dimensions is constructed
with the aid of two mutually orthogonal null vectors which naturally
arise in a generalized spacetime superalgebra. An obstacle encountered
in an attempt to extend this result to beyond 14 dimensions is
described. A null reduction of the $(11,3)$ model is shown to yield the
known super Yang-Mills model in $(10,2)$ dimensions. An $(8,8)$
supersymmetric super Yang-Mills system in $(3,3)$ dimensions is obtained
by an ordinary dimensional reduction of the $(11,3)$ model, and it is
suggested there may exist a superbrane with $(3,3)$ dimensional
worldvolume propagating in $(11,3)$ dimensions.

\end{titlepage}

\section{ $(3,3)$ Superbrane in $(11,3)$ Dimensions?}

There are several reasons for exploring supersymmetry in higher than
eleven dimensions. Many of our motivations for considering supersymmetry
in $(10,2)$ dimensions in particular have already been discussed in
\cite{ns}, where additional references can be found. Here, we will
especially emphasize the fact that a) the F-theory considerations
\cite{vafa} have shown the power of $(10,2)$ dimensional framework for
unifying a large class of string vacua in a nontrivial way, b) the (2,1)
string approach to $ M$-theory \cite{km} has also pointed at a $(10,2)$
dimensional target space, and c) in an algebraic approach to unifying the
perturbative and nonperturbative superstring states \cite{ht,t,by}
evidence has been put forward for a $(10,2)$ dimensional structure
\cite{b1}. Later, possible existence of hidden symmetries descending
from (11,2) dimensions was proposed \cite{b2}.

Very recently \cite{bk2}, it has been suggested that the fundamental
supersymmetric theory may admit as many as 11 spacelike and 3 timelike
dimensions \cite{b3}. This observation has motivated us to look for an
extension of the work presented in \cite{ns}, in search of a
supersymmetric field theory in $(11,3)$ dimensions, the simplest one
being super Yang-Mills theory. Interestingly enough, we find that the
construction of \cite{ns} generalizes naturally to $(11,3)$ dimensions,
while an extension beyond $(11,3)$ dimensions runs into an obstacle. 

The existence of the model constructed here seems to require two
mutually orthogonal null vectors. These are essential for a null
reduction to $(10,2)$ dimensions, yielding the results of \cite{ns}, or
to the $(9,1)$ dimensional super Yang- Mills theory by a further null
reduction. It is also possible to obtain an $(8,8)$ supersymmetric
Yang-Mills system in $(3,3)$ dimensions by means of an ordinary dimensional
reduction. The result contains the two mutually orthogonal null vectors
inherited from $(11,3)$ dimensions.

The $(11,3)\rightarrow (3,3)$ reduction is similar to the $(10,2)$
$\rightarrow$ (2,2) reduction, where the resulting theory in $(2,2)$
dimensions is relevant to the target space of the $(2,1)$ string of
\cite{km}. This fact and the close relationship observed between the
$(8+n,n)$ theories and their $(n,n)$ reductions for $n=1,2,3$ prompts us
to suggest that there may exist a superbrane with a $(3,3)$ dimensional
worldvolume propagating in $(11,3)$ dimensions. In view of the intricate
and fascinating results that continue to emerge in six dimensional
physics, it is conceivable that this proposal finds a realization. This,
in turn, may play a significant role in unifying a large class of
duality symmetries in an interesting way. In particular, the fact that
the isometry group $SO(3,3)\sim SL(4,R)$ is the conformal group for a
$(2,2)$ dimensional world may be of relevance in this picture.

While we are not aware of any literature on supersymmetric field
theories in $(3,3)$ dimensions, we note ref. \cite{dks} where a
generalized self-duality condition on the Yang-Mills curvature in
$(3,3)$ dimensions was shown to produce the KP equation. 

We now turn to the description of the super Yang-Mills system in
$(11,3)$ dimensions. As a prelude to doing so, we first describe a
general class of superalgebras in $(8+n,n)$ dimensions, then reviewing
briefly the $(10,2)$ results. After describing the $(11,3)$ model and
its dimensional reductions, we will explain the obstacle to the
construction in higher than 14 dimensions in the final section.

\section{$(1,0)$ Superalgebras in $(8+n,n)$ Dimensions}

Let us consider a spacetime superalgebra which contains a single
Majorana-Weyl spinor generator $Q_\a$~which has $2^{n+3}$ real
components
\footnote{One can equally well work with pseudo Majorana-Weyl spinors.
It can be easily verified that this choice does not alter the form of the
algebra. Note also that the symmetry property of the matrices
$\c^{m_1\cdots \m_r}C^{-1}$ always repeats itself for $r~{\rm mod}~4$.
See \cite{kt} for further details.} 
and all possible bosonic generators that can occur in their
anticommutation relation. Hence there are
$2^{n+2}\left(2^{n+3}+1\right)$ bosonic generators, in addition to the
Lorentz generators. The chirality and symmetry properties of the
$\c$-matrices are identical for $n~{\rm mod}~4$. Hence, we have the
following algebras 
\bea
&&n=0~{\rm mod}~4:   \nn\\
&&\{ Q_\a, Q_\b\} = \eta_{\a\b}~Z 
    + (\c^{\m_1\cdots \m_4})_{\a\b}~Z_{\m_1\cdots \m_4}
    +\cdots +(\c^{\m_1\cdots \m_{n+4}})_{\a\b}~Z_{\m_1\cdots \m_{n+4}}\ ,\la{a0} \\
&&\nn\\
&&n=1~{\rm mod}~4:   \nn\\
&&\{Q_\a, Q_\b\} = (\c^\m)_{\a\b}~P_\m  
+ (\c^{\m_1\cdots \m_5})_{\a\b}~Z_{\m_1\cdots \m_5}
+\cdots +(\c^{\m_1\cdots \m_{n+4}})_{\a\b}~Z_{\m_1\cdots \m_{n+4}}\ , \la{a1}\\
&&\nn\\
&&n=2~{\rm mod}~4:   \nn\\
&&\{ Q_\a, Q_\b\} = (\c^{\m\n})_{\a\b}~Z_{\m\n} 
    + (\c^{\m_1\cdots \m_6})_{\a\b}~Z_{\m_1\cdots \m_6}
    +\cdots +(\c^{\m_1\cdots \m_{n+4}})_{\a\b}~Z_{\m_1\cdots \m_{n+4}}\ ,\la{a2}\\ 
&&\nn\\
&&n=3~{\rm mod}~4:   \nn\\
&&\{Q_\a, Q_\b\} = (\c^{\m\n\r})_{\a\b}~Z_{\m\n\r}
+ (\c^{\m_1\cdots \m_7})_{\a\b}~Z_{\m_1\cdots \m_7}
+\cdots +(\c^{\m_1\cdots \m_{n+4}})_{\a\b}~Z_{\m_1\cdots \m_{n+4}}\ , \la{a3}
\eea  
where $Z_{m_1\cdots m_p}$ are the bosonic $p$-form generators, all of
which commute with each other and with $P_\m$. The commutators involving
the Lorentz generators are the usual ones. These algebras are mapped
into each other by a dimensional reduction on a (1,1) dimensional
internal space, followed by a chiral truncation. 

The $\c$-matrices $\c^{\m_1\cdots \m_p}$ are actually the chirally projected
$(\c^{\m_1\cdots\m_p}{C}^{-1})$, where $C$ is the charge conjugation
matrix in $(8+n,n)$ dimensions. The $\eta$-matrix occurring in (1) is the chiral
projection of $C$. In this chiral notation, the spinor index takes the
values $\a=1,...,2^{n+3}$. In a given dimension, all possible symmetric
$\c$- matrices that survive the chiral projection occur on the right
hand side, and in all cases the maximal rank $\c$-matrix has a definite
duality property. Hence, we take the generator $Z_{\m_1\cdots \m_{n+4}}$
to be self-dual. Taking this into account, one can easily confirm that
the r.h.s. of the above algebras span the full symmetric space of 
relevant dimension. 

Notice that a vector momentum operator $P_\m$ occurs only in the case of
$n=1~mod~4$. Since all the $p$-form generators $Z_{\m_1\cdots \m_p}$
correspond to charges that can be carried by $p$-branes, one should also
add $Z_\m$ to the algebra to allow string charges. While this may be
redefined away by shifting $P_\m$, there are some global subtleties in
doing so, and they have an interesting role to play in the description
of string winding states \cite{t}.

It is worth mentioning that the case of (1,0) algebra in $(9,1)$
dimensions admits a non Abelian extension which involves a super 1-form
and a super 5-form generator \cite{es}. Whether the $(8+n,n)$ algebras with
$n>1$ admit a similar non Abelian extension is an interesting open
question.

Given the $mod~4$ repetitive character of the above algebras, there is a
sense in which the $(11,3)$ dimensions is a natural maximum dimension,
namely it is the last member of the first quartet. We will consider the
case of (12,4) later, but to keep the discussion and calculations
tractable, let us consider the cases of $n=0,1,2,3$, and keep the lowest
rank bosonic generators: 
\bea
&(8,0):\qquad \{ Q_\a, Q_\b\} = \eta_{\a\b}~Z \ , \la{b0} \\
&\nn\\
&(9,1):\qquad \{Q_\a, Q_\b\} = (\c^\m)_{\a\b}~P_\m \ , \la{b1}\\
&\nn\\
&(10,2):\qquad \{ Q_\a, Q_\b\} =(\c^{\m\n})_{\a\b}~Z_{\m\n} \ ,\la{b2}\\ 
&\nn\\
&(11,3):\qquad \{Q_\a, Q_\b\} = (\c^{\m\n\r})_{\a\b}~Z_{\m\n\r} \ . \la{b3}
\eea 

Note that in each one of these cases there is only one more
$Z$-generator, which is of rank $n+4$ and self-dual. This fact will be
significant later, when we discuss the obstacle to constructing
super Yang-Mills in higher than 14 dimensions (see section 6). 

The (8,0) algebra, though interesting in its own right and may as well
have certain applications, it can not provide a basis for an acceptable
spacetime since it is timeless. The $(9,1)$ algebra is the well known
Poincar\'e superalgebra. In the case of $(10,2)$ dimensions, while there
is no vector momentum generator, there is a way to introduce it by
introducing a constant (null) vector $n_\m$ into the algebra as follows
\cite{ns,b2}
\be
(10,2):\qquad \{ Q_\a, Q_\b\} =(\c^{\m\n})_{\a\b}~P_\m~n_\n \ .\la{alg2}\\ 
\ee
The constancy of the vector $n_\m$ should be considered as a special
case of a more general situation where $n_\m$ is another momentum
generator. Indeed, in \cite{bk2,bk1} just such a scenario has been
advocated, and an interesting two-particle interpretation has been put
forward. It is beyond the scope of this paper to review these ideas in
any detail here.

In the next section, we will recall the results of \cite{ns}, after
which we will present the generalization to $(11,3)$ dimensions.

\section{ Recalling Super Yang-Mills in $(10,2)$ Dimensions}

The Yang-Mills equations of motion are given by \cite{ns}
\bea
&&\c^\m D_\m \l=0\ , \la{f1}\\
&&D^\m F_{\m[\r} n_{\s]} - \ft12 \bar\l\c_{\r\s} \l = 0 \ ,\la{f2}
\eea
where the fields are Lie algebra valued and in the adjoint
representation of the Yang-Mills gauge group, and $D_\m
\l=\p_\m\l+[A_\m,\l]$. Due to the symmetry of $\c^{\m\n}C^{-1}$, the
last term in \eq{f2} involves a commutator of the Lie algebra generators. 
In addition to the manifest Yang-Mills gauge symmetry, these equations are 
invariant under the supersymmetry transformations \cite{ns}
\bea
\d_Q A_\m &=& \bar\e\c_\m\l \ , \la{t1} \\
\d_Q  \l &=& -\ft14 \c^{\m\n\r} \e F_{\m\n} n_\r \ , \la{t2}
\eea
and the extra bosonic local gauge transformation \cite{ns}
\be
\d_\Omega A_\m = \Omega~n_\m \ , \quad\quad \d_\Omega \l = 0\ , \la{to}\la{ot12}
\ee
provided that the following conditions hold \cite{ns}
\bea
 n^\m D_\m \l &=& 0  \ ,   \la{c1}\\
 n^\m \c_\m \l &=&0 \ ,    \la{c2}\\
 n^\m F_{\m\n} &=& 0 \ ,   \la{c3}\\
 n^\m n_\m &=& 0 \ ,       \la{c4}\\
 n^\m D_\m \Omega &=& 0 \ ,\la{c5}
\eea
One can check that the field equations as well as the constraints are
invariant under supersymmetry as well as extra gauge transformations.

Finally, we recall that the commutator of two supersymmetry
transformations closes on shell, and yields a generalized translation,
the usual Yang-Mills gauge transformation and an extra gauge
transformation with parameters $\xi^\mu$, $\Lambda$, $\Omega$,
respectively, as follows:
\be
[ \d_Q(\e_1), \d_Q(\e_2) ] = \d_\xi +\d_\Lambda + \d_\Omega\ ,
\la{closure}
\ee
where the composite parameters are given by \cite{ns}
\bea
\xi^\m &=&  \bar\e_2 \c^{\m\n}\e_1\ n_\n\ , \la{xi}\\
\Lambda &=& -\xi^\m\ A_\m\ , \la{l}\\
\Omega &=& \ft12 \bar\e_2 \c^{\m\n}\e_1\ F_{\m\n}\ .  \la{o}
\eea
Note that the global part of the algebra \eq{closure} is given by
\eq{alg2}.

The closure of the supersymmetry algebra on the fermion requires the
constraints \eq{c1} and \eq{c2}, while the supersymmetry and $\Omega$-
symmetry of the field equations and constraints require the remaining
constraints as well \cite{ns}. A superspace formulation of this model,
as well as its null reductions to $(9,1)$ and (2,2) can be found in
\cite{ns}. As we will see in the next section, most of these results
have a natural generalization to $(11,3)$ dimensions. 

\section{Super Yang-Mills in $(11,3)$ Dimensions}

We begin by introducing the momentum generator to the algebra \eq{b3},
by making use of a constant tensor $v_{\m\n}$ as follows:
\be 
\{Q_\a, Q_\b\} = (\c^{\m\n\r})_{\a\b}~P_\m~v_{\n\r} \ . \la{alg3}
\ee
The next step is to postulate the supersymmetry transformation rules
which make use of $v_{\m\n}$. The strategy is then to obtain the field
equations, and any additional constraints by demanding the closure of
these transformation rules. At the end the (extra) gauge and
supersymmetry of all the resulting equations must be established. In
what follows, we will first present the results that emerge out of this
procedure. Later, we will explain the step by step derivation of these
results.

The super Yang-Mills equations take the form
\bea
&&\c^\m D_\m \l=0\ , \la{nf1}\\
&&D^\s F_{\s[\m} v_{\n\r]} + \ft1{12}\bar\l \c_{\m\n\r} \l = 0 \ .\la{nf2}
\eea
In addition to the manifest Yang-Mills gauge symmetry, these equations
are invariant under the supersymmetry transformations
\bea
\d_Q A_\m &=& \bar\e\c_\m\l \ , \la{nt1} \\
\d_Q  \l &=& -\ft14 \c^{\m\n\r\s} \e F_{\m\n} v_{\r\s} \ , \la{nt2}
\eea
and the extra bosonic local gauge transformation
\be
\d_\Omega A_\m = -v_{\m\n}~\Omega^\n \ ,\quad\quad 
\d_\Omega \l = 0\ , \la{not}
\ee
provided that the following conditions hold:
\bea
 v_\m{}^\n D_\n \l &=& 0  \ , \la{nc1}\\
 v_{\m\n} \c^\n \l &=&0 \ , \la{nc2}\\
 v_\m{}^\n F_{\n\r} &=& 0 \ , \la{nc3}\\
 v_\m{}^\r v_{\r\n} &=& 0 \ , \la{nc}\\
 v_{[\m\n} v_{\r\s]} &=& 0 \ , \la{nc4}\\
 v_\m{}^\r v_\n{}^\s D_\r \Omega_\s &=& 0  \ , \la{nc5}\\
 v^{\m\n} D_\m \Omega_\n &=& 0\ . \la{nc6}
\eea

The commutator of two supersymmetry transformations closes on shell, and
yields a generalized translation, the usual Yang-Mills gauge
transformation and an extra gauge transformation with parameters
$\xi^\mu$, $\Lambda$, $\Omega^\m$, respectively, as follows:
\be
[ \d_Q(\e_1), \d_Q(\e_2) ] = \d_\xi +\d_\Lambda + \d_\Omega\ ,
\la{nclosure}
\ee
where the composite parameters are given by
\bea
\xi^\m &=&  \bar\e_2 \c^{\m\n\r}\e_1\ v_{\n\r}\ , \la{nxi}\\
\Lambda &=& -\xi^\m\ A_\m\ , \la{nl}\\
\Omega^\m &=& \ft12 \bar\e_2 \c^{\m\n\r}\e_1\ F_{\n\r}\ .  \la{no}
\eea
The global part of the algebra \eq{nclosure} indeed agrees with
\eq{alg3}. Note the symmetry between the parameters $\xi^\m$ and
$\Omega^\m$. The former involves a contraction with $v_{\m\n}$, and the
latter one with $F_{\m\n}$.

The derivation of these results proceeds as follows. First, it is
easy to check that the closure on the gauge field requires an additional
local gauge transformation \eq{not} with the composite parameter
\eq{no}. Next, one checks the closure on the gauge fermion. In doing so, the following 
Fierz-rearrangement formula is useful:
\be
\e_{[1}\bar\e_{2]} = \ft1{64}
	\left( \ft1{3!}~\bar\e_2 \c^{\m\n\r}\e_1~\c_{\m\n\r} 
	+\ft1{7!2}~\bar\e_2\c^{\m_1\cdots \m_7}\e_1~\c_{\m_1\cdots\m_7}\right)\ ,
	\la{f}
\ee
where $\e_1$ and $\e_2$ are Majorana-Weyl spinors of the same chirality.
Using this formula, and after a little bit of algebra, one finds that:

(a)  The closure on the gauge fermion holds provided that the fermionic field
equation \eq{nf1}, along with the constraints \eq{nc1} and \eq{nc2} are
satisfied.

(b) The supersymmetry of the constraint \eq{nc1} requires the constraint
\eq{nc3}, and a further variation of this constraint does not yield
new information.

(c) The supersymmetry of the constraint \eq{nc2} requires the further 
constraints \eq{nc4} and \eq{nc5}. 

(d) The equations of motion \eq{nf1} and \eq{nf2} transform into each other under 
supersymmetry. This can be shown with the use the constraints \eq{nc2} and
\eq{nc3}.

(e) Finally the invariance of the full system, i.e. equations of motion
and constraints, under the extra gauge transformation \eq{not} has to be
verified. The invariance of the fermionic field equation \eq{nf1}, as
well as the constraints \eq{nc1} and \eq{nc2} do not impose new
conditions. However, the invariance of the constraint \eq{nc3} imposes
the condition \eq{nc5}, and the invariance of the bosonic field equation
\eq{nf2} imposes the condition \eq{nc6} on the parameter $\Omega^\m$.
Both of these conditions are gauge invariant.

In summary, equations \eq{alg3}-\eq{no} form a consistent and closed
system of supersymmetric, Yang-Mills gauge and $\Omega$-gauge invariant
equations. The similarity of these equations to the corresponding ones in
$(10,2)$ dimensions is evident. One expects, therefore, a natural
reduction of these equations to those in $(10,2)$ dimensions. This will
indeed turn out to be the case, as we will see in the next section.

The important next step is to establish that the constant tensor
$v_{\m\n}$ satisfying the conditions \eq{nc4} and \eq{nc} actually
exists. Fortunately this is the case, and we have the solution
\be 
v_{\m\n}=m_{[\m} n_{\n]}\ , \la{vmn} 
\ee 
where $m_\m$ and $n_\n$ are mutually orthogonal null vectors, i.e. they
satisfy 
\be
m_\m m^\m =0\ ,\quad\quad n_\m n^\m =0\ , \quad\quad m^\m n_\m = 0\ . \la{mn} 
\ee 

Given the signature of the 14-dimensional spacetime, finding two
mutually orthogonal null vectors, of course, does not present a problem.
Indeed, this solution suggests a that an {\it ordinary} dimensional
reduction to $(9,1)$ dimensions should yield the usual super
Yang-Mills system. In the next section we will show that
this is indeed the case.

\section{Dimensional Reductions to $(10,2)$ and $(3,3)$}

The simplest way to show that the ordinary dimensional reduction of the
$(11,3)$ system to $(9,1)$ dimensions yields the usual super Yang-Mills
equations is to establish that the full system of equations in $(10,3)$
dimensions reduce to those in $(10,2)$ dimensions. Since the latter have
already been shown to reduce to the usual super Yang-Mills equations in
$(9,1)$ dimensions \cite{ns}, we need not repeat the second step of
dimensional reduction.

In this section we shall use hats for the fields and indices in $(11,3)$,
to distinguish them from the unhatted ones in $(10,2)$. The coordinates
are $~(x^\m, x^{13}, x^{14})$~ and the metric is
$~(\Hat\eta_{\hat\m\hat\n} ) = (\eta_{\m\n},+,-)$, where $x^\m$ are the
coordinates of the $(10,2)$ dimensional space with metric
$\eta_{\m\n}=\hbox{diag.}~(-,+,\cdots,+,-)$. It is convenient to define
the coordinates $x^\pm\equiv(x^{11}\pm x^{12})/{\sqrt2}$. 

The $(11,3)$ $\c$-matrices satisfy $~\{\Hat\c_{\hat\m}, \Hat\c_{\hat\n}\}
= 2 \Hat\eta_{\hat\m\hat\n}$. A convenient choice for the $\c$-matrix is
\be
\Hat\c^{\hat\m} = \cases{\hbox{$\Hat\c^\m = \c^\m \otimes \s_3 ~~,$} \cr
\hbox{$\Hat\c^{13} = I\otimes \s_1 ~~,$} \cr
\hbox{$\Hat\c^{14} = I \otimes i\s_2 ~~.$} \cr }
\ee
Here $I$ is the $64\times 64$ unit matrix and the $\s$'s are the Pauli
matrices and $\c^\m$ are the $64\times 64$ Dirac $\c$-matrices in $(10,2)$
dimensions. The charge conjugation matrix and the chirality operators can 
defined as
\be
\Hat C= C_{12}\otimes i\s_2\ , \quad\quad 
\Hat \c_{15}=\c_{13}\otimes \s_3\ , \la{cc}
\ee 
where $C_{12}$ is the antisymmetric charge conjugation matrix in $(10,2)$
such that $\c^\m (C_{12})^{-1}$ is antisymmetric, and $\c^{13}$ is the
chirality operator in $(10,2)$ which squares to one. The mutually
orthogonal null vectors $m_\m$ and $n_\m$ are taken
to be
\be
\left( \Hat m_{\hat \m}\right) = ({\vec 0}, 1, 1 )\ , \quad\quad
\left( \Hat n_{\hat \m}\right) = (n_\m, 0, 0 )\ , \la{mnc}
\ee
where $n_\m$ is a null vector in $(10,2)$ dimensions.

With these choices, the $v$-tensor has the components $v_{\m\n}=0$,
$v_{-\m}=0$ and $v_{+\m}=n_\m$.  Without making any assumption on the 
$x^\pm$ dependence of the fields and parameters, the bosonic constraint
\eq{nc3} reduces to 
\bea
&& F_{-\m}=0\ ,\qquad F_{-+}=0\ ,\qquad n^\m F_{+\m}=0\ , \la{fc}\\
&& n^\m F_{\m\n}=0\ . \la{rc3}
\eea
Similarly the gauge transformations \eq{not}, including the Yang-Mills 
gauge transformations become
\bea
\d A_- &=& D_-\L\ , \qquad \d A_+ = D_+\L -n_\m \Omega^\m\ ,  \la{gt1}\\
\d A_\m &=& D_\m \L + n_\m \Omega^+\ , \la{gt2}
\eea
with the constraint \eq{nc5} on the parameter $\Omega^\m$ reducing to
\bea
&&D_-\Omega^+ = 0\ ,\qquad D_-\left( n_\r\Omega^\r\right) = 0\ ,
\qquad  n^\m D_\m \left( n_\r \Omega^\r\right) = 0\ , \la{oc1}\\
&& n^\m D_\m \Omega^+ = 0\ . \la{rc5}
\eea

The reduction of the fermionic constraints \eq{nc1} and \eq{nc2} is also
straightforward. The latter one gives $\c^+ \l= 0$, where $\c^+ =
I\otimes (\s_1+ i\s_2) /{\sqrt 2}$. Together with the $(11,3)$ chirality
condition, this implies that $\Hat\lambda $ can be written as
\be
\Hat \lambda = \pmatrix{ \lambda \cr 0 \cr }\ , \qquad\qquad
\c_{13}\lambda=\lambda\ . \la{rl}
\ee 
The fermionic constraint \eq{nc2} now reduces to
\be
n_\m \c^\m \l = 0\ , \la{rc2}
\ee
and the constraint  \eq{nc1} gives
\bea
D_- \l &=& 0\ , \la{rd}\\
n^\m D_\m \l &=& 0 \la{rc1}\ .
\eea
There remains the reduction of the field equations and the supersymmetry
transformation rules. It is easily seen that equations of motion reduce to
\bea
&& \c^\m D_\m \l =0 \ , \la{rs}\\
&& D^\m F_{\m[\r} n_{\s]} - \ft1{8} \bar\l \c_{\r\s} \l= 0 \ . \la{nfe}
\eea
and the supersymmetry transformation take the form
\bea
&&\d_Q A_- = 0\ , \quad\quad \d_Q A_+= \bar\eta \l\ , \la{ntr1}\\
&&\d_Q A_\m = \bar\epsilon \c_\m \l\ , \la{ntr2}\\
&&\d_Q \l = \c^{\m\n\r}\e F_{\m\n} n_\r\ , \la{ntr3}
\eea
where we have used the notation
\be
\Hat \epsilon = \pmatrix{ \eta \cr \epsilon \cr }\ . \la{de}
\ee 
$\eta$ is a Majorana-Weyl spinor in $(10,2)$ with chirality opposite to
that of $\epsilon$. The $\eta$-transformation of $A_+$ was omitted in
\cite{ns} but this is inconsequential since $A_+$ can be gauged away by
an $\Omega$-transformation.

Up to global issues which may involve large gauge transformations and
nontrivial topologies, equations \eq{fc}-\eq{de} describe the super
Yang-Mills system in $(10,2)$ dimensions, in which an arbitrary
dependence on $x^+$ is introduced but the derivative $\p_+$ does not
occur. Integrating over $x^+$ then yields precisely the super Yang-Mills
equations of \cite{ns}. Though the details may differ slightly, this
phenomenon is similar in essence to the null reduction from $(10,2)$ to
$(9,1)$ discussed in \cite{ns}. Thus, a two-step reduction to $(9,1)$
dimensions is expected to yield a doubly affinized version of the usual
super Yang-Mill theory in $(9,1)$ dimensions. Applied to the present
case, the single step reduction argument goes as follows. 
 
Firstly, the field $A_+$ can be gauged away by using the last two
constraints in \eq{fc} and the second gauge transformation in \eq{gt1}.
In doing so, the last two equations in \eq{oc1} needs to be taken into
account. Second, the field $A_-$ can be gauged away by using the first
two constraints in \eq{fc} and the first gauge transformation in
\eq{gt1}. The surviving fields are $A_\m(x^+, {\vec x})$ and 
$\l(x^+, {\vec x})$ obeying the constraints \eq{rc3}, \eq{rc2} and \eq{rc1}. 
Here ${\vec x}$ represents the $(10,2)$ coordinates. The surviving
symmetries are the rigid supersymmetry transformations \eq{ntr2} and
\eq{ntr3}, and the gauge transformations \eq{gt2} with parameters
$\L(x^+, {\vec x})$ and $\Omega^+(x^+, {\vec x})$ subject to the condition
\eq{rc5}. With the identification $\Omega^+ \equiv \Omega$, and suitable
rescalings of the fields and parameters, this system is exactly the
super Yang-Mills system of \cite{ns} as summarized in eqs.
\eq{f1}-\eq{o}, with the additional and arbitrary $x^+$ dependence
inherited from $(11,3)$ dimensions.

To conclude this section, we describe the ordinary dimensional reduction
to $(3,3)$ dimensions, in which case the extra coordinates become part of
the resulting spacetime. To begin with, let us label the coordinates as
\be
x^{\hat\m}=(x^\mu, x^i)\ ,\quad \m=0,1,11,...,14\ ,\quad  i=2,...,9\ . 
\ee
Thus the signature of the $(3,3)$ spacetime is $(-++-+-)$, and the
internal space is Euclidean. The ordinary dimensional reduction from
$(11,3)$ to $(3,3)$ is achieved simply by setting
\be
\p_i=0\ ,\quad v_{ij}=0\ ,\quad v_{i\m}=0\ , \la{con}
\ee
and using the Dirac matrices $\Hat\c_\m = \c_\m \otimes I$ and $\Hat
\c_i = \C_7\otimes \c_i $, where $\c_\m$ and $\c_i$ are the $8\times 8$
$SO(3,3)$ and $16\times 16$ $SO(8)$ Dirac matrices, respectively, and
$\C_7$ is the chirality operator in $(3,3)$ dimensions and $I$ is the unit
matrix. 

The resulting fields are $(A_\mu, \phi^i, \l^{A}, \l^{\dot A})$, where
the eight scalars are defined as $A_i\equiv \phi_i$, and the spinors are
Majorana-Weyl in $(3,3)$ dimensions, with the indices $A,\dot A=1,...,8$
labelling the left and right handed spinors of the internal $SO(8)$ group. 
It is a straightforward matter to apply \eq{con} to all the equations
from \eq{nf1} to \eq{no}. The resulting system is clearly of the same
form as the one in $(11,3)$ dimensions in that it contains the tensor
$v_{\m\n}$ in a similar fashion. There are, of course, the contributions
of the eight scalars $A_i$ to the equations which are easily obtained
from the $(11,3)$ equations. 

The superalgebra in $(3,3)$ dimensions that underlies this model is an
(8,8) type superalgebra with 8 left-handed and 8 right-handed
Majorana-Weyl spinor generators. Using the chiral notation in which the
lower and upper SO(3,3) spinor indices refer to left and right handed
projections, the superalgebra takes the form
\bea 
\{Q_\a^A, Q_\b^B\} &=& \d^{AB}\, (\c^{\m\n\r})_{\a\b}~P_\m~v_{\n\r} \ ,\la{1}\\
\{Q_\a^A, Q^{\b\dot B} \} &=& 0\ ,\la{2}\\
\{Q^{\a\dot A}, Q^{\b\dot B}\} &=& \d^{\dot A\dot B}\, (\c_{\m\n\r})^{\a\b}~P^\m~v^{\n\r} 
\ . \la{3}
\eea
Although $(p,q)$ type chiral version of this algebra perfectly exists,
just as in $(11,3)$ dimensions, the field theoretic realization obtained
by ordinary dimensional reduction is vectorlike, unless one imposes the
self-duality condition
\be
p^{\m_1} v^{\m_2\m_3}= 
	\ft1{3!} \epsilon^{\m_1\cdots \m_6}\,p_{\m_4} v_{\m_5\m_6}\ , \la{dc}
\ee
in which case the algebra becomes $(8,0)$ with the anticommutation relation 
\eq{1}.

\section{Beyond 14 Dimensions and Comments}

The $(11,3)$ model suggests a
generalization to $(8+n,n)$ dimensions for all values of $n$. One way to
proceed is to keep the lowest rank bosonic generators, namely $P_\m$ for
$n=1~{\rm mod}~4$; $Z_{\m\n}$ for $n=2~{\rm mod}~4$ and $Z_{\m\n\r}$ for
$n=3~{\rm mod}~4$. However, supersymmetry transformations fail to close
on the gauge fermion for $n>3$, due to the appearance of new and
unwanted contributions to the relevant Fierz identity. Another
approach would be to introduce higher rank generators into the algebra
that take the form
\be
v_{\m_1\cdots \m_p} = n_{[\m_1} \cdots n_{\m_p]}\ ,
\ee
for one or more suitable values of $p$. If such generators are
introduced in addition to the lowest rank generators mentioned above, the
problem with the closure of the algebra on the gauge fermion persists. A
third approach would be to introduce the $v$-tensor for a particular
$p$-form generator. The simplest case to consider is the superalgebra in
$(12,4)$ dimensions:
\be
\{ Q_\a, Q_\b\} = (\c^{\m\n\r\s})_{\a\b}~P_\m v_{\n\r\s} \ . \la{16d}
\ee
Postulating the obvious analogs of the supersymmetry transformation
rules \eq{nt1} and \eq{nt2} for this case, one finds that the algebra
indeed closes modulo constraints analogous to \eq{nc1}-\eq{nc6}, and the
fermionic field equation \eq{nf1}. However, a supersymmetric variation
of the fermionic field equation $\c^\m D_\m \l=0$ does not yield a
bosonic field equation, and therefore the system fails to be
supersymmetric. The problem arises in the variation of the gauge field
in the covariant derivative. It requires the Fierz rearrangement formula
\be
\e_{[1}\bar\e_{2]} = \ft1{128}
	\left( \bar\e_2 \e_1+ \ft1{4!}~\bar\e_2 \c^{\m_1\cdots \m_4}
	\e_1~\c_{\m_1\cdots \m_4}
	+\ft1{8!2}~\bar\e_2\c^{\m_1\cdots \m_8}\e_1~\c_{\m_1\cdots \m_8}\right)\ ,
	\la{ff}
\ee
where $\e_1$ and $\e_2$ are Majorana-Weyl spinors of the same chirality.
The last term is acceptable because $\c^\r \c_{\m_1\cdots \m_8}\c_\r
=0$. The second term is also acceptable because it produces the
$\bar\l\c_{\m_1\cdots \m_4}\l$ term in the Yang-Mills equation. However
the first term in \eq{ff} gives an unwanted contribution which can not
be interpreted as part of the Yang-Mills field equation\footnote{As this
problem does not arise in the Abelian case, the construction works for
super Maxwell for any $n$. However, we do not consider this to be
interesting since it is a free theory.}. The reason why the construction
works for $n=1,2,3$ is that the Fierz rearrangement formula in those
cases gives only two terms, one of which is harmless due to the formula
$\c^\r \c_{\m_1\cdots \m_{n+4}}\c_\r =0$ in $(2n+8)$-dimensions, and the
other gives rise to a fermionic bilinear term in the Yang-Mills
equation.

The obstacle mentioned above arises for all $n>3$, since the unwanted
lower rank $\c$-matrix contributions to the relevant Fierz rearrangement
formula would set in. While this is a tentative analysis and in
principle new structures may be introduced to the algebra to avoid the
obstacle, it is nonetheless interesting to see that there is something
special about the lowest triplet of theories in the $(8+n,n)$
dimensional class, namely the ones corresponding to $n=1,2,3$, at least
within the current framework. 
 
To conclude, we turn to the case of $(11,3)$ dimensions and note that 
the algebra \eq{alg3} can be generalized to 
\be 
\{Q_\a, Q_\b\} =
(\c^{\m\n\r})_{\a\b}~P_{1\m}~P_{2\n}~P_{3\r} \ , \la{alg4} 
\ee where
$P_{1\m}$, $P_{2\n}$ and $P_{3\r}$ are to be considered on equal footing
as momentum generators, just as it has been suggested in
\cite{b1,b2,bk2,bk1} for the case of $(10,2)$ dimensions where two such
momenta arise. Recently an interesting two-particle interpretation has
been given for that case \cite{bk2}. This approach is indeed promising
because among its premises is a manifestly $SO(10,2)$ invariant action,
albeit with the introduction of bi-local fields. The arguments of
\cite{bk2} suggest, however, that bi-local fields need not necessarily
suffer from the old problems. A suitable application of these ideas to
the present case would presumably involve tri-local fields, and possibly
a notion of a new kind of triality symmetry. It would be very
interesting to see if such a picture might emerge within the framework
of a $(3,3)$ superbrane propagating in $(11,3)$ dimensions mentioned in
the introduction.

\vfill\eject

\ed